# Fire in SRRN: Next-Gen 3D Temperature Field Reconstruction Technology

Shenxiang Feng,[1] Xiaojian Hao,[1,*] Xiaodong Huang,[1] Pan Pei,[1] Tong Wei,[1] Chenyang Xu[1]

[1]*State Key Laboratory of Dynamic Measurement Technology, North University of China, Taiyuan 030051 Shanxi, China*
**haoxiaojian@nuc.edu.cn*

**Abstract:** In aerospace and energy engineering, accurate 3D combustion field temperature measurement is critical. The resolution of traditional methods based on algebraic iteration is limited by the initial voxel division. This study introduces a novel method for reconstructing three-dimensional temperature fields using the Spatial Radiation Representation Network (SRRN). This method utilizes the flame thermal radiation characteristics and differentiable rendering in graphics, and combines it with a multi-layer perceptron to achieve a functional representation of the flame temperature field. The effectiveness of SRRN is evaluated through simulated temperature field reconstruction experiments with different levels of complexity. The maximum root mean square error is 10.17, which proves the robustness of the algorithm to Gaussian noise and salt-and-pepper noise. We conducted a butane flame temperature field reconstruction experiment, and the maximum relative error between the reconstruction result and the thermocouple measurement value was 4.86%, confirming that the algorithm can achieve accurate reconstruction.

## 1. Introduction

Combustion field temperature measurement is essential in aerospace, energy engineering, and other fields[1]. Accurate reconstruction of the three-dimensional temperature field of the combustion flame not only helps to study the combustion process but also has great potential in improving combustion efficiency, assessing high-temperature damage, and controlling pollution emissions [2,3]. Current temperature measurement methods are mainly divided into contact and non-contact. Contact temperature measurement technology has high accuracy and low cost, but it can usually only measure point temperature, and intrusive detection may affect the temperature field. Non-contact temperature measurement technology has attracted more attention from researchers due to its advantages, such as wide temperature measurement range, fast response speed, and strong anti-interference ability[4].

Non-contact temperature measurement methods include acoustic methods and optical methods. The acoustic method mainly reconstructs the temperature field based on the relationship between temperature and sound wave speed. Qian Kong et al. proposed a 3D temperature field reconstruction algorithm based on polynomial reproductive radial basis function approximation and truncated generalized singular value decomposition to solve the problem of sound waves in non-uniform temperature fields—the refraction effect in [5]. However, the acoustic thermometry method is easily affected by background noise. At the same time, to improve the reconstruction accuracy, more acoustic generators and receivers need to be added, resulting in increased costs.

The optical temperature measurement method is another critical non-contact temperature measurement method, including laser spectroscopy and radiation spectroscopy. Although laser spectroscopy methods such as laser interference holography [6], planar laser-induced fluorescence (PLIF) [7], and coherent anti-Stokes Raman scattering spectroscopy (CARS) have achieved accurate temperature measurements[8,9], However, its reliance on high-precision optical equipment and its high-cost limit its application in extreme industrial environments. In contrast, radiation spectroscopy, especially radiation spectrum imaging in the infrared and visible bands, has been widely used in surface temperature measurements of high-temperature targets. Temperature measurement based on radiation spectroscopy has accurately obtained surface temperature information of high-temperature objects [10], but there are still challenges in reconstructing the three-dimensional temperature field. J. Floyd et al. proposed chemiluminescence computational tomography technology, which uses two-dimensional projections (camera images) of the flame in different directions to reconstruct the 3D chemiluminescence field of the turbulent flame [11]. Although this method classifies the three-dimensional reconstruction of flames as the inverse process of projection to three dimensions for the first time, it still uses discrete voxels and algebraic iterative reconstruction algorithms for solutions. The reconstruction speed is slow, and the resolution is low. Ying Jin et al. proposed a three-dimensional flame chemiluminescence tomography reconstruction system based on a convolutional neural network model, using pre-collected data to train the network to reconstruct three-dimensional flame distribution rapidly. However, the 3D ground truth used in the training of this algorithm is reconstructed by the ART algorithm, resulting in the accuracy of the three-dimensional flame distribution output by the CNN being lower than that of the ART algorithm [12]. Cai W et al. tried to use transfer learning and semi-supervised learning algorithms to achieve 3D flame reconstruction based on small-scale training samples, which reduced the difficulty of constructing training sets to a certain extent [13], but this method still did not solve the 3D ground truth proposed by ART The algorithm reconstructs this vital issue.

Recent research has shown that NeRF outperform computational imaging on non-cooperative objects without texture, reflective and refractive surfaces [14]. Inspired by this, we propose a three-dimensional temperature field reconstruction algorithm based on a spatial radiation representation neural network (SRRN), which performs 3D reconstruction based on the 2D projection of the flame in different directions. The reconstructed temperature field is reprojected through differentiable rendering technology, and the mean square error (MSE) between the reprojection and the actual projection is used as the loss function to train the model. This reconstruction method converts the three-dimensional temperature field reconstruction problem into a function fitting problem. Use a multilayer perceptron(MLP) to fit the mapping relationship between spatial coordinates and point temperature. Therefore, this method can output a three-dimensional temperature field with any resolution and temperature information at any position in the combustion field.

The subsequent parts of the article are organized as follows: Section 2 explains the basic principles of radiation thermometry and the thermo-optical radiation model; Section 3 introduces in depth the algorithm details of the spatial radiation representation network SRRN; Section 4 uses numerical simulation experiments and three-dimensional flame temperature The actual reconstruction of the field is performed to verify the effectiveness of the proposed method; finally, the article provides a summary and outlook.

## 2. Principle of radiation thermometry and thermal-optical radiation model

In order to use neural networks to represent the three-dimensional temperature field, we need first to understand the principle of radiation thermometry and the thermal and optical radiation transfer model of flames[15,16].

### 2.1 Principle of radiation thermometry

Flame is a highly complex gas-phase chemical reaction system, and its internal radiation is mainly generated by reaction products generated by combustion and fuel particles that have not been completely burned. Under thermal excitation, these particles will transition from a high-energy state to a low-energy state, releasing photons and producing flame radiation [17]. Radiation thermometry technology is based on the inherent relationship between the radiation intensity emitted by an object at high temperatures and its temperature[18]. The temperature information of the object is calculated by measuring the intensity of radiation energy emitted by the object. From a technical perspective, radiation temperature measurement methods are divided into the following three categories: total radiation method based on radiant energy within the entire wavelength range, radiation temperature measurement method based on a single wavelength, and temperature measurement method based on the ratio of radiant energy of two different wavelengths. This article uses the single-wavelength radiation thermometry method. According to Planck's law, the relationship between the radiation intensity of wavelength \lambda at any point inside the flame and its absolute temperature is as shown in Equation (1)

$$I(\lambda,T) = \varepsilon(\lambda) \frac{2\pi hc^2}{\lambda^5 \left(e^{hc/\lambda kT} - 1\right)} \tag{1}$$

where $\lambda$ is the wavelength, $c$ is the speed of light, and $n$ is the spectral emissivity of the non-black body, which is determined through calibration experiments. $h$ is Planck's constant, $k$ is Boltzmann's constant. The integral formula of the radiation generated by the flame passing through the camera lens on the imaging plane is shown in Equation (2)

$$L(p) = \int_0^\infty \omega(x) I(x) dx \tag{2}$$

where $L(p)$ is the radiation intensity integral received by the pixel $p$ on the imaging plane, $I(x)$ and $\omega(x)$ are the radiation intensity and corresponding weight at the midpoint $x$ in the space. After photoelectric conversion inside the camera, the radiation intensity integral $L(p)$ is converted into a grayscale value $G(p)$. Through the blackbody furnace calibration test, the relationship between the gray value $G(p)$ and the temperature $T$ can be determined, as shown in Equation (3)

$$T = A \times G(p) + B \tag{3}$$

where A and B are constants. Based on the above principles, the two-dimensional temperature projection can be calculated from the flame projection image collected by the camera. Our previous work improved the imaging grayscale by doping the flame with K+ elements [14-16]. Therefore, in this paper, a narrow-band filter with a central wavelength of 768nm was installed in front of the lens to measure the flame temperature.

*2.2 Flame thermal and optical radiation model*

Traditional volume reconstruction techniques generally reconstruct the three-dimensional temperature field of the flame by deconvolving a limited number of projections[19]. In order to establish the mathematical relationship between the reconstructed temperature field and its projection, the reconstruction target domain needs to be divided into discrete sets of voxels before reconstruction. Then, an imaging model is established to determine the mapping relationship between voxels and pixels. However, dividing the target domain into a fixed number of voxels directly limits the resolution of the reconstructed three-dimensional temperature field. If we want to improve the reconstruction accuracy, we must use a more complex projection method to determine the voxel corresponding to each pixel, increasing the difficulty of reconstruction and the computational cost.

In order to solve the dependence on voxels in traditional reconstruction methods, we established a new flame thermal and optical radiation model by analyzing the thermal and optical radiation process of the flame. The thermal and optical radiation of the flame is divided into three stages: thermal radiation excitation of the internal nodes of the flame, radiation transfer from the nodes to the camera, and internal imaging of the camera. First, the internal nodes of the flame are thermally excited to produce radiation, which reaches the imaging plane after attenuation through transmission media such as the interior of the flame, air, and camera optics. Since the attenuation effect of air and optical devices on flame radiation is weak within short distances, the effect of attenuation is not considered in this imaging model.

The small hole imaging model (the pinhole camera model) is often used in computer vision to describe the two-dimensional projection of an object captured in three-dimensional space. In order to facilitate the reconstruction of the three-dimensional temperature field, we establish the flame thermal and optical radiation model of Equation (4), simplify the camera into a viewpoint and an imaging plane, and use a forward imaging model to facilitate the definition of light. The projected temperature value of each pixel is the weighted integral of the temperature value of the light at each point on the target domain.

$$T(p) = \int_n^f t(s) ds \tag{4}$$

In equation (4), $n$ and $f$ represent the distance from the sampling point on the light to the imaging plane, $p$ represent the temperature value at the sampling point s, and $T(p)$ is the temperature value of the pixel p. The mapping relationship between pixel p and sampling point s is determined by the internal and external parameter matrix of the camera, as shown in Equation (5)

$$Z_c \begin{bmatrix} u \\ v \\ 1 \end{bmatrix} = \begin{bmatrix} f_x & 0 & c_x & 0 \\ 0 & f_y & c_y & 0 \\ 0 & 0 & 1 & 0 \end{bmatrix} \begin{bmatrix} R & T \\ \vec{0} & 1 \end{bmatrix} \begin{bmatrix} X_w \\ Y_w \\ Z_w \\ 1 \end{bmatrix} \tag{5}$$

where $(u, v)$ is the position of pixel p in the pixel coordinate system, $Z_c$ is the distance from the sampling point to the imaging plane in the space under the camera coordinate system, $f_x$, $f_y$, $c_x$, $c_y$ are the internal parameters of the camera, R and T are the rotation matrices of the camera respectively. and translation matrix, $(X_w, Y_w, Z_w)$ are the world coordinates of the spatial sampling point s.

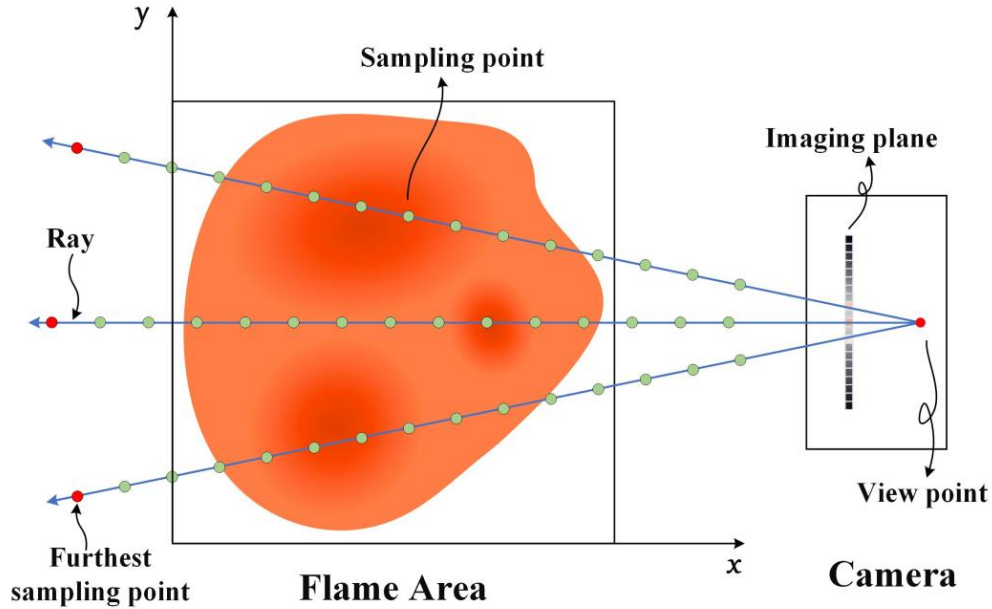

Fig. 1. 2D principle of ray sampling. Establish world coordinates to determine camera coordinates and spatial sampling point coordinates. Simplifying the camera into a viewpoint and an imaging plane, starting from the viewpoint, rays are emitted into space toward the pixels. Sampling equidistantly along the ray. Set the same maximum sampling distance for all rays and sample between the camera viewpoint and the farthest sampling point.

Based on the above model, we can perform uniform sampling, as shown in Fig. 1, along the light ray, thereby obtaining the sampling point coordinates at any position in the reconstruction target domain corresponding to each pixel in the camera.

## 3. SRRN and its key technologies

In the method of using supervised learning to reconstruct the temperature field, obtaining the ground truth of the three-dimensional temperature field is a big challenge. As a new perspective synthesis technology, NeRF can also be used for three-dimensional field reconstruction, and this technology is very effective in reconstructing transparent objects such as glass. In previous research, we tried to use NeRF to reconstruct the three-dimensional shape of a stable candle flame and achieved results that traditional 3D reconstruction algorithms could not achieve. Based on NeRF's idea of using neural networks to represent three-dimensional scenes, we further explored the application of implicit representation in three-dimensional temperature field reconstruction.

NeRF can achieve good reconstruction results for transparent and translucent objects similar to flames because it follows the physical laws of light scattering, reflection, and diffraction. As long as the angle at which image data is collected is sufficient, the sampling light generated based on the projection data from these perspectives can cover the entire target domain, and the neural network will represent the target domain more accurately. Usually, NeRF requires dozens or even hundreds of images for model training. If the collection perspectives are too sparse, the model will be difficult to converge. Drastic changes often accompany combustion reactions, and sufficient projection data cannot be obtained unless a large number of cameras are used to capture them simultaneously. Since flame, as a self-illuminating translucent medium, has a relatively simple composition compared to the complex scenes targeted by NeRF, the three-dimensional temperature field reconstruction under sparse viewing angles can be achieved by improving the sampling method of the flame area.

In this study, we propose a thermo-optic radiation rendering technology based on the flame thermo-optic radiation model proposed in Section 2.2 and the differentiable rendering technology in computer graphics, and at the same time, construct a fully connected neural network for representing the three-dimensional temperature field. In order to make it easier for readers to understand the algorithm principle of the spatial radiation representation network, we first introduce the two key technologies used by SRRN in section 3.1 and then introduce the overall workflow and architecture of the model in detail in Section 3.2.

*3.1 Key technologies*

The SRRN model uses spatial position coding technology and thermo-optical radiation rendering technology. The spatial position coding technology improves the model's ability to represent the flame boundary and the rapidly changing area inside the flame. The thermo-optical radiation rendering technology is used to calculate the emission of each pixel. The projection value produced by the sampled ray. These three key technologies will be introduced below.

**Spatial position encoding technology**: As a continuous medium, flame often has significant changes in the temperature field due to uneven chemical reactions. As a general function approximator, the neural network is not good at fitting high-frequency functions. For three-dimensional temperature field reconstruction, it is manifested in insufficient fitting ability for areas with severe temperature changes. The spatial position coding technology maps the input spatial coordinates to a high-dimensional space, thereby converting the high-frequency function fitting problem into the low-frequency function fitting that the neural network is good at, ultimately improving the MLP's ability to represent the rapidly changing areas of the three-dimensional temperature field and at the same time Speed up model training. The spatial position encoding function is shown in Equation (6)

$$\Gamma(s) = \left(\sin(2^0 \pi s), \cos(2^0 \pi s), \cdots, \sin(2^{L-1} \pi s), \cos(2^{L-1} \pi s)\right) \quad (6)$$

where $s$ is the spatial coordinate of the sampling point, and $L$ is the encoded vector dimension. In this study we set $L=5$ to encode each dimension of spatial coordinates, so the fully connected network input vector size is 33.

**Thermal radiation rendering technology**: Differentiable rendering technology has been a research focus in the intersection of computer graphics and deep learning in recent years. Its core idea is to model the traditional rendering process as a differentiable function. For any parameter in the scene (such as the object's geometry, surface material, light source properties, etc.), its derivative or gradient to the final rendered image can be calculated, making it possible to use neural networks to represent three-dimensional fields. Establishing a correct rendering formula for three-dimensional temperature field reconstruction is the key to obtaining accurate two-dimensional projections. This study proposes a thermo-optical radiation rendering formula based on differentiable rendering technology and a thermo-optical radiation model. First, a ray is emitted from the camera viewpoint toward the pixel, with the direction shown in equation (7)

$$\vec{r}(t) = \vec{o} + l\vec{d} \quad (7)$$

where $\vec{r}$ represents the light emitted by the pixel, $\vec{o}$ is the world coordinate of the camera viewpoint, $\vec{d}$ is the direction of the emitted light, and $l$ represents the distance of the light forward. Due to the characteristics of the semi-transparent medium of the flame itself, the radiation intensity integral on each pixel is shown in Equation (8)

$$T(r) = \int_r t(s) ds \quad (8)$$

where **r** represents the light emitted by the pixel, s is the coordinate of each point on the ray **r**, $t(s)$ represents the temperature value at the sampling point s, $T(r)$ is the temperature value corresponding to the pixel emitting ray **r**. Through this formula we can obtain the reprojection result of the temperature field to facilitate comparison with the original projection.

### 3.2 Pipeline of SRRN

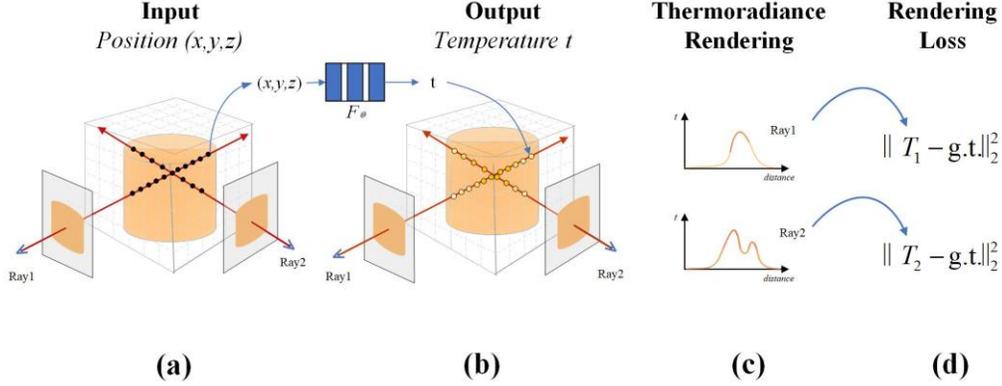

Fig. 2. Pipeline of SRRN. Figure (a) shows the sampling point generation. Figure (b) shows the calculation of sampling point temperature. Figures (c) and (d) show the reprojection along the ray and the calculation of the loss function, respectively.

**Generate sampling rays and sampling points** First, generate a dataset of rays based on all two-dimensional projections and the corresponding camera pose. Since the flame temperature field is a scalar field with a single parameter, uniform sampling points can be generated along the light ray. In this study, we uniformly generate 45 sampling points with a sampling interval of 1 along the ray, enough to cover the entire reconstruction target domain. The sampling formula is as follows

$$s_i \sim u[s_n + \frac{i-1}{N}(s_f - s_n), s_n + \frac{i}{N}(s_f - s_n)] \tag{9}$$

where $s_n$ and $s_f$ represent the nearest and farthest sampling points on the light, $N$ is the total number of sampling points, and the absolute sampling range $[s_n, s_f]$ and sampling step $s_{step}$ are determined.

**Sampling point position encoding** The flame temperature field is a continuously changing area. In order to improve the network's ability to represent the rapidly changing temperature area, formula (4) is used to encode the position of each dimension of all sampling points in the space and map them to high dimensional space. In this study, we set the vector dimension of the encoding function to 5. After encoding, the original spatial coordinates and the encoded coordinates are spliced, and the coordinate dimension of the input network is 33.

**Pixel reprojection** Each pixel corresponds to a sampling ray. All sampling points on this ray are position-encoded and then input into the MLP to obtain the temperature value output of each sampling point. The reprojection value of this pixel is calculated according to Equation (6) of the rendering technology. In this study, we take the mean square error of the reprojected value and the actual projected value as the loss function. The loss function of each batch is calculated as shown in Equation (10)

$$\mathcal{L} = \sum_{r \in \mathcal{R}} \left[ \| T_p(r) - T(r) \|_2^2 \right] \tag{10}$$

where $\mathcal{R}$ is the sampling light of each batch, $T_p(r)$ is the projection value of the sampling light, and $T(r)$ is the temperature value of the pixel that generates the sampling light. After obtaining the reprojection error for each ray, we update the model parameters using the Adam optimizer, with the initial learning rate of the optimizer set to 10e-4 and the weight set to 0.95.

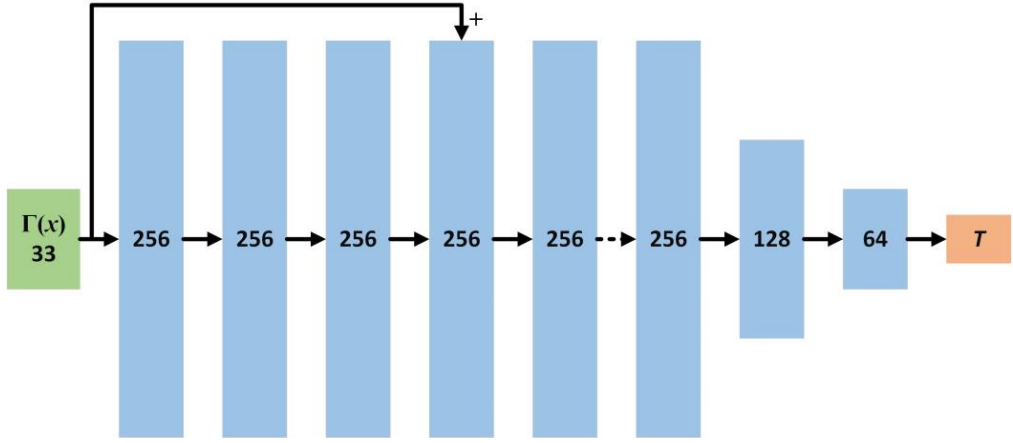

Fig. 3. Architecture of SRRN.

The SRRN network structure is shown in Figure 3. The input vector is green, the middle hidden layer is blue, the output vector is orange, and the number within each block indicates the dimension of the vector. All layers are fully connected; the black arrow indicates the layer with ReLU activation, and the output layer is corrected using ReLU to ensure that the output result is non-negative. The coordinates after position encoding are input into the network through 6 fully connected layers with ReLU as the activation function, and the width of each hidden layer is 256. Next, the downsampling layers with widths of 128 and 64 are connected to reduce the dimension, and finally, the temperature value of the point is output. In order to alleviate the vanishing and exploding gradient problems in deep networks, we use a residual connection to concatenate the input with the vectors of the fourth layer. The "+" in the figure indicates vector concatenation.

## 4. Numerical verification and experiments

In order to evaluate the performance of SRRN in three-dimensional temperature field reconstruction, we conducted numerical simulation experiments and butane flame temperature field reconstruction experiments.

The purpose of carrying out numerical simulation experiments in this study is to verify the feasibility and noise immunity of the algorithm and to evaluate the temperature field reconstruction effect of different complexities and the impact of noise on the reconstruction results. In order to facilitate the evaluation of the three-dimensional field reconstruction effect, the reconstruction target domain is divided into voxels, SRRN is used to output the temperature value of each voxel, and the root mean square error(RMSE) from the original temperature value is calculated using the following formula

$$RMSE = \sqrt{\frac{1}{N}\sum_{i}\left\|T_i - \hat{T}_i\right\|^2} \tag{11}$$

where $T_i$ and $\hat{T}_i$ represent the reconstructed value and original temperature value corresponding to voxel $i$ respectively. $N$ is the number of discrete voxels.

In order to further verify the algorithm's reconstruction effect on the actual flame temperature field, we carried out a three-dimensional temperature field reconstruction experiment of butane flame. The reconstruction accuracy of the temperature field was evaluated by taking points inside the reconstructed butane flame and comparing them with thermocouple measurements.

4.1 Numerical simulation experiment

In this study, to verify the algorithm's feasibility and ability to reconstruct complex temperature fields, we used single fireball, dual fireball, and three fireballs to simulate temperature fields of different complexity. Each fireball uses a spherical Gaussian distribution to simulate the temperature distribution of the flame, as shown in Equation (12)

$$T = T_{max} \cdot e^{-K} \tag{12}$$

where $T_{max}$ is the maximum value of the temperature field, which we set to 1000K, K is used to control the temperature distribution, as shown in Equation (13)

$$K = \frac{(x-x_0)^2 + (y-y_0)^2 + (z-z_0)^2}{R^2} \tag{13}$$

where $(x_0, y_0, z_0)$ is the spherical center coordinate of a single spherical temperature field, and R is the radius of the fireball.

Three temperature field average simulations were reconstructed using 12 cameras, which were evenly arranged at 30-degree intervals at the same horizontal height, as shown in Figure 4. It can be seen from the figure that the temperature field of a single Gaussian distribution is smooth and continuous inside. The temperature fields of double Gaussian distribution and triple Gaussian distribution not only have a smooth area inside the temperature field but also have a rapidly changing area at the junction of the two Gaussian distributions.

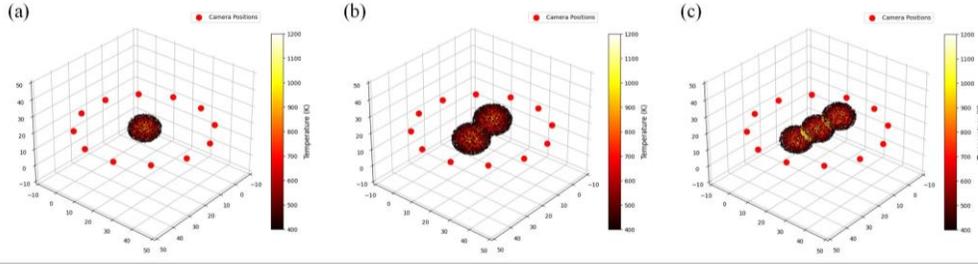

Fig. 4. Distribution of simulated temperature field and camera. Simulate the temperature field and camera distribution. The red dot in the figure represents the camera position, Figure (a) represents the temperature field of a single fireball, and Figure (b) and (c) represent the temperature fields of double fireballs and three fireballs.

The training errors in the reconstruction process of the three types of temperature fields are shown in Figure 5. Among them, the single fireball temperature field model converges the fastest due to its relatively simple structure, and the average MSE loss of each epoch after convergence is the smallest. As the temperature field complexity increases, the number of iterations required for model convergence increases, and the average MSE loss of each epoch also increases, but the difference is negligible. It can be seen from the figure that the three types of temperature fields have reached convergence after 20 epochs of training. There are abnormal loss values during the reconstruction process of the three fireball temperature fields, possibly due to the significant learning rate in the early model training stage. However, it can be seen from the loss value after the model converges that as the training proceeds, the Adam optimizer gradually reduces the learning rate, allowing the model to converge typically.

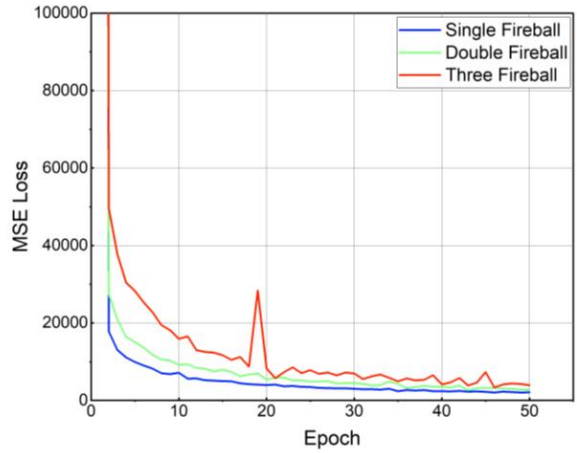

Fig. 5. Reconstruction reprojection errors. The three curves in the figure respectively represent the average MSE of each epoch during the training process of single fireball, double fireball, and triple fireball.

In Figure 6, we show the three-dimensional slices of three temperature field reconstruction results and three original temperature field slices at z=-14, z=0, and z=16 heights, two-dimensional slices of the reconstructed temperature field, and their relative error maps. It can be seen from the three-dimensional slice that SRRN has an excellent ability to represent various temperature fields, and the non-flame area and the flame area have apparent boundaries. As is shown in the two-dimensional relative error distribution diagram that the errors are mainly concentrated in the boundary part of the flame, while the errors inside and outside the flame are minimal. Calculate the root mean square errors of the three reconstructed temperature fields according to Equation (7). The RMSE of the single fireball, double fireball, and three fireball temperature fields are 4.62, 10.11, and 10.17, respectively. Among them, the single-fireball temperature field reconstruction error is the smallest, and the double and three fireball reconstruction errors are close.

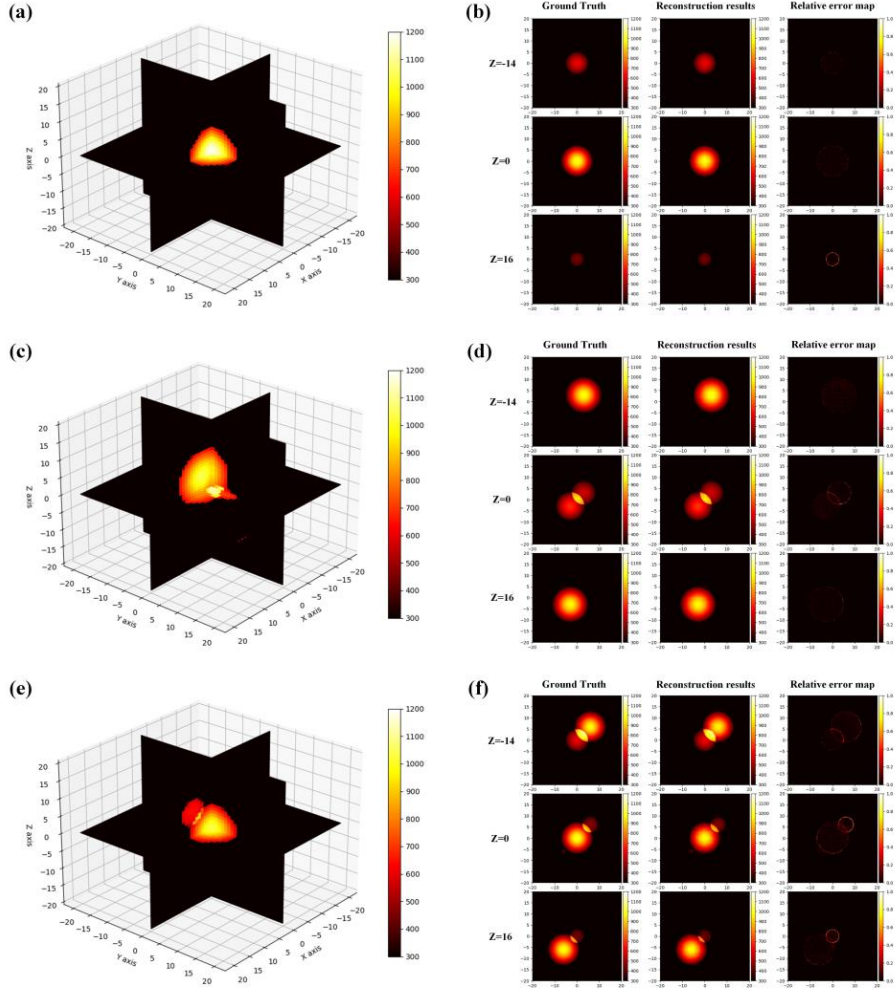

Fig. 6. Temperature field reconstruction slices. Figures (a) and (b) show the three-dimensional slices of a single fireball temperature field, and the two-dimensional slices and relative error distribution at three heights z=-14, 0, and 16. Figures (c) and (d) show the three-dimensional slices of the temperature field of the double fireball, and the two-dimensional slices and relative error distribution at three heights z=-14, 0, and 16. Figures (e) and (f) represent three-dimensional slices of the temperature field of three fireballs, and two-dimensional slices and relative error distribution at three heights z=-14, 0, and 16.

## 4.2 Anti-noise experiment

In the three-dimensional temperature field measurement, the quality of the flame image collected by the camera is affected by factors such as soot particles produced by incomplete combustion of the fuel, environmental soot, and light refraction due to heat in the air, thus reducing the reconstruction effect of the three-dimensional temperature field of the flame. In this study, we added Gaussian noise and salt-and-pepper noise to the two-dimensional projection used for reconstruction to verify the anti-noise ability of the algorithm. Gaussian noise simulates ambient light, and salt-and-pepper noise simulates smoke particles.

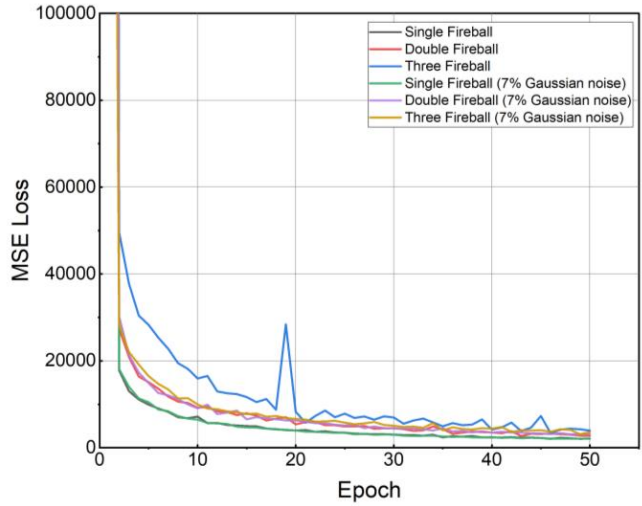

Fig. 7. Reconstruction rejection errors with 7% Gaussian noise. The figure lists three types of temperature field reprojection errors in the absence and presence of noise.

In this experiment, Gaussian noise with an intensity of 7% is first added to the two-dimensional projections in all directions. Figure 7 shows the loss change curve during SRRN model training after adding 7% Gaussian noise and the loss change during model training without noise. It can be seen from the figure that the model training loss curve after adding Gaussian noise is highly coincident with the loss curve without noise. After adding Gaussian noise, we calculated the RMSE between reconstructing temperature fields of different complexities and the original temperature field. The reconstruction error of the single fireball temperature field was 6.18, and the errors of the double fireball temperature field and the three fireball temperature field are 10.86 and 13.09 respectively. Figure 8 further visually displays the reconstruction results of the three-dimensional temperature field under 7% Gaussian noise in the form of slices. It can be seen from the three-dimensional slices that the reconstructed temperature field has clear boundaries and is relatively smooth internally. The relative error plot of the two-dimensional slice shows that the reconstruction error is mainly concentrated at the flame boundary.

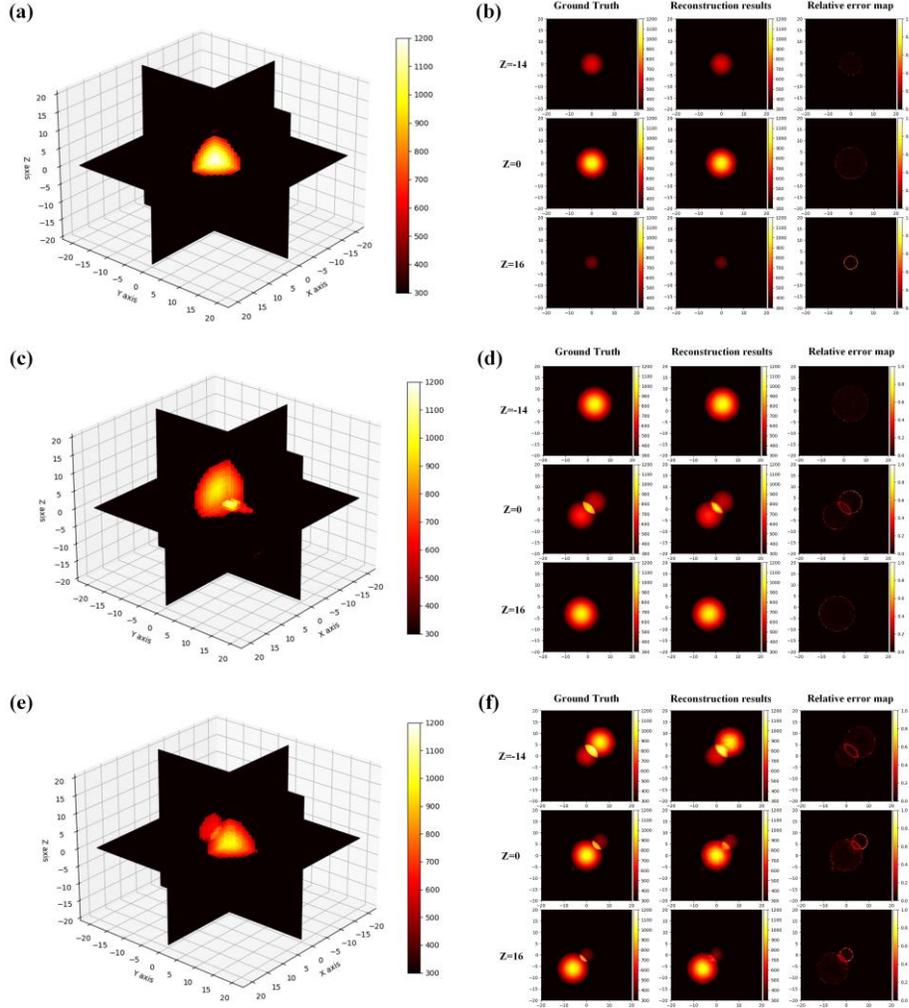

Fig. 8. Reconstruction of temperature field slices with 7% Gaussian noise. Figures (a) and (b) show the three-dimensional slices of a single fireball temperature field with 7% Gaussian noise, and the two-dimensional slices and relative error distribution at three heights z=-14, 0, and 16. Figures (c) and (d) show the three-dimensional slices of the temperature field of a double fireball with 7% Gaussian noise, and the two-dimensional slices and relative error distribution at three heights z=-14, 0, and 16. Figures (e) and (f) represent the three-dimensional slices of the temperature field of three fireballs with 7% Gaussian noise, and the two-dimensional slices and relative error distribution at three heights z=-14, 0, and 16.

In order to further study the impact of Gaussian noise on the three-dimensional temperature field reconstruction, we added Gaussian noise with an intensity of 15% to the projection data. It can be seen from the loss change curve in Figure 9 that the loss value of model convergence after adding 15% Gaussian noise is slightly worse compared to the case without noise. Figure 10 shows the reconstruction results in the form of slices. It can be seen from the reconstructed slices that the reconstruction effect is close to the reconstruction result under 7% Gaussian noise, and the error is concentrated at the flame boundary position where the temperature changes drastically. The root mean square errors of the reconstruction of the three types of temperature fields are as follows. The reconstruction error of the single fireball temperature field is 6.06. And the errors of the double fireball temperature field and the three fireball temperature field are 10.40 and 12.84, respectively. Compared with 7% Gaussian noise, The three-dimensional temperature field reconstruction result has no significant change in error. Therefore, we believe that Gaussian noise has little impact on the accuracy of the SRRN model in representing the three-dimensional temperature field, and the model has strong resistance to Gaussian noise.

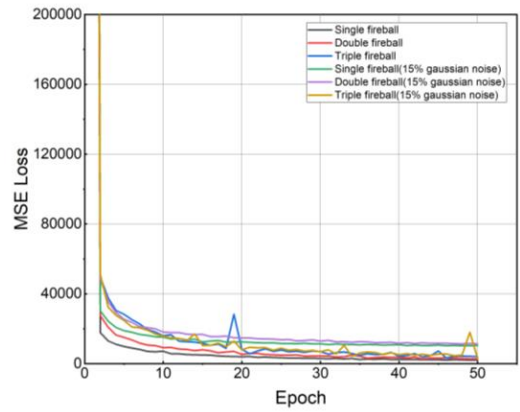

Fig. 9. Reconstruction reprojection errors with 15% Gaussian noise. The figure lists three types of temperature field reprojection errors in the absence and presence of noise.

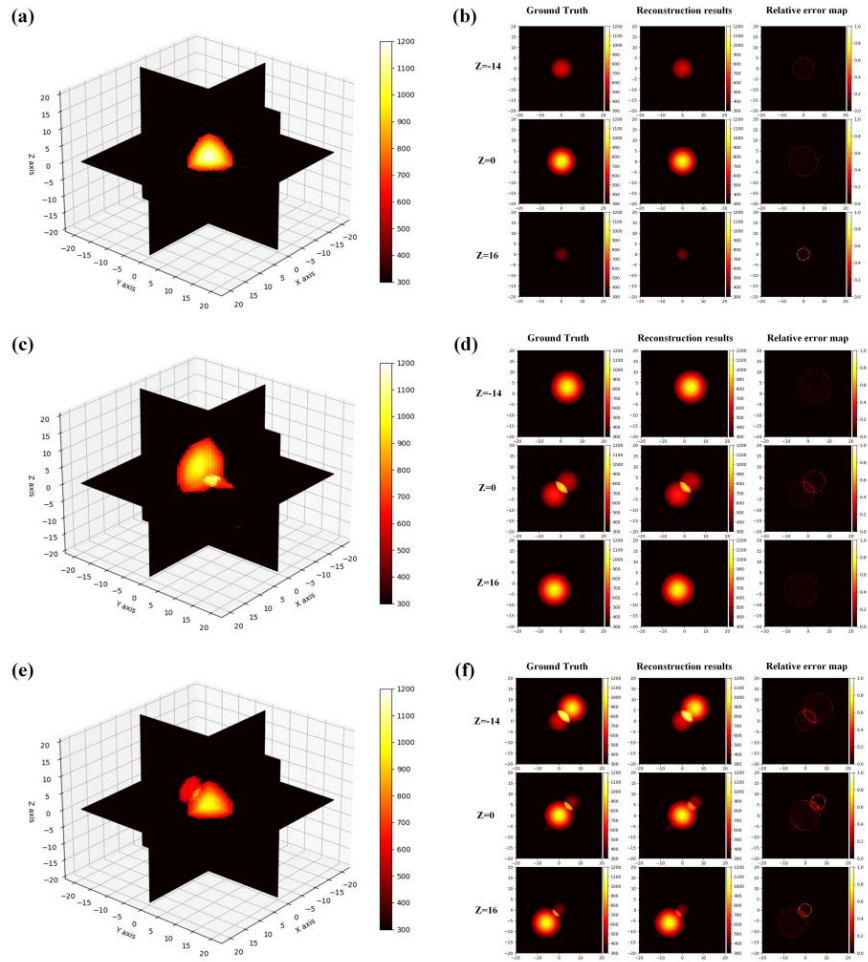

Fig. 10. Reconstruction of temperature field slices with 15% Gaussian noise. Figures (a) and (b) show the three-dimensional slices of a single fireball temperature field with 15% Gaussian noise, and the two-dimensional slices and relative error distribution at three heights z=-14, 0, and 16. Figures (c) and (d) show the three-dimensional slices of the temperature field of a double fireball with 15% Gaussian noise, and the two-dimensional slices and relative error distribution at three heights z=-14, 0, and 16. Figures (e) and (f) represent the three-dimensional slices of the temperature field of three fireballs with 15% Gaussian noise, and the two-dimensional slices and relative error distribution at three heights z=-14, 0, and 16.

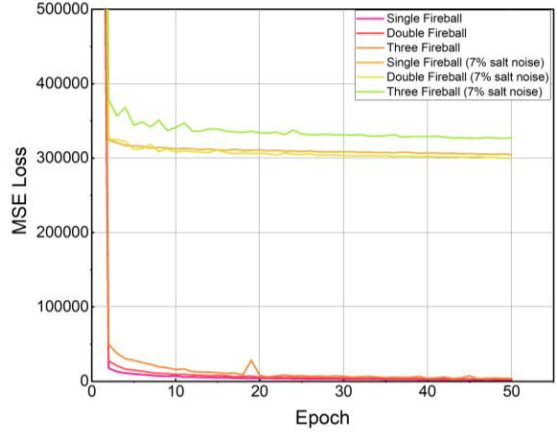

Fig. 11. Reconstruction reprojection errors with 7% salt noise. The figure lists three types of temperature field reprojection errors in the absence and presence of noise.

Like Gaussian noise, we first add salt and pepper noise with an intensity of 7% to the two-dimensional projection to simulate the impact of factors such as smoke and dust on the reconstruction of the temperature field. Figure 11 shows the SRRN model training loss curve after adding salt and pepper noise. It can be seen from the figure that the influence of salt and pepper noise on the reconstruction results is evident. Because salt and pepper noise directly causes missing parts of projection data and errors in projection values, according to the light sampling method of SRRN, errors in a single projection value will directly affect the temperature values of all sampling points on the corresponding light, thereby reducing the reconstruction accuracy. Gaussian noise has little impact on reconstruction accuracy because it changes the size of the projection value, and when this change is averaged over all sampling points on the light, its impact is almost negligible. The impact of salt and pepper noise on the reconstruction results can be seen from the temperature field slice in Figure 12. Although the reconstructed temperature field still has a relatively obvious boundary due to the missing part of the projection, compared with the noise-free reconstruction result, there are more apparent boundaries at the boundary. Obvious burrs. There are also significant errors in the reconstruction results inside the temperature field. According to equation (7), the RMSE of the temperature field reconstruction for single, double, and three fireballs are 10.43, 19.41, and 26.06, respectively.

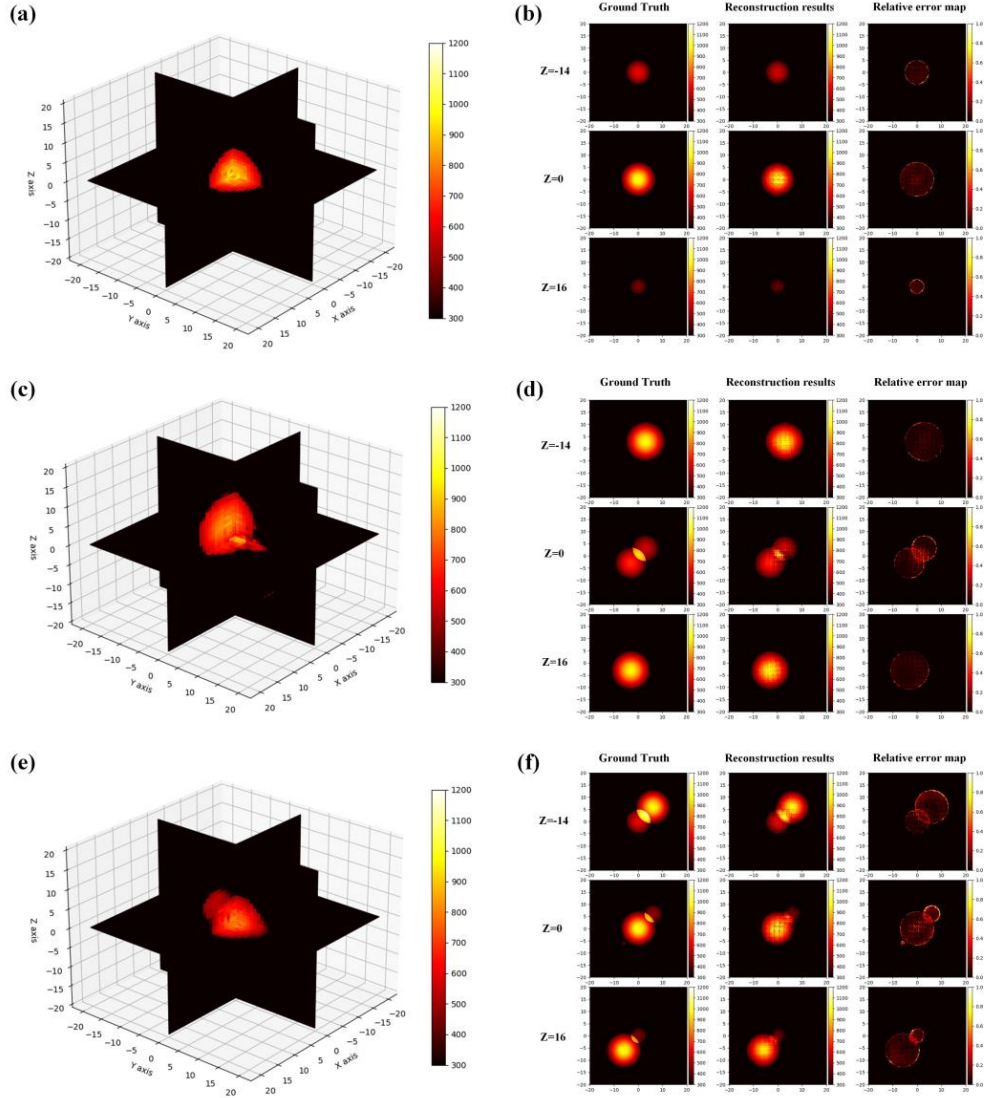

Fig. 12. Reconstruction of temperature field slices with 7% salt noise. Figures (a) and (b) show the three-dimensional slices of a single fireball temperature field with 7% salt noise, and the two-dimensional slices and relative error distribution at three heights z=-14, 0, and 16. Figures (c) and (d) show the three-dimensional slices of the temperature field of a double fireball with 7% salt noise, and the two-dimensional slices and relative error distribution at three heights z=-14, 0, and 16. Figures (e) and (f) represent the three-dimensional slices of the temperature field of three fireballs with 7% salt noise, and the two-dimensional slices and relative error distribution at three heights z=-14, 0, and 16.

*4.3 Flame reconstruction experiment*

In order to further verify the algorithm's reconstruction effect on the flame temperature field, we conducted a three-dimensional temperature field reconstruction experiment on a butane flame. The maximum temperature of the butane flame can reach 1400K [16]. In this study, the accuracy of the three-dimensional temperature field reconstruction was finally determined by comparing the errors between the measured values of thermocouples at multiple sampling points inside the temperature field and the reconstructed temperature values.

Since the jet flame is in a rapidly changing state due to the disturbance of the airflow, we use the synchronous control system, as shown in Fig. 13, to ensure that all cameras collect the projection data of the flame at the same time. The synchronous control system mainly uses a signal generator to send acquisition signals to all cameras simultaneously, and a high-speed data acquisition card transmits the image data captured by 12 cameras to the workstation for processing. A total of 12 CMOS cameras were used in this experiment. The distance between all cameras and the flame was 50cm. They were evenly arranged with the target flame as the center and 30 degrees apart. The internal and external parameters of the camera were obtained using COLMAP calibration [17]. In addition, each camera lens is equipped with a narrow-band filter with a central wavelength of 768nm and a bandwidth of 10nm.

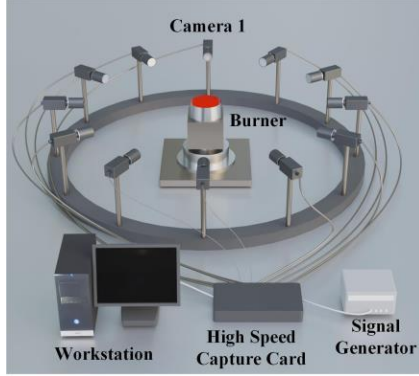

Fig. 13. Schematic diagram of butane flame reconstruction device. The 12 cameras in the figure are evenly arranged with the flame as the center and a radius of 50cm, spaced at 30 ° intervals. The signal generator and acquisition card achieve multi-channel data synchronous acquisition.

Before reconstructing the three-dimensional temperature field of the flame, it is necessary to determine the relationship between grayscale and temperature values in the 768nm band through camera temperature calibration. We use a medium-temperature blackbody furnace for calibration. The calibration temperature range is 873.15K to 1273.15K, and data is collected every 25K. After processing, the relationship between temperature and grayscale is obtained, as shown in Equation 14, and the corresponding fitting curve is shown in Figure 14.

$$T = -16387.7\exp\left(-\frac{G}{1.63}\right) - 257.9\exp\left(-\frac{G}{12.57}\right) - 261.7\exp\left(-\frac{G}{137.06}\right) + 1326.4 \quad (14)$$

Where $G$ is the corresponding gray value of the pixel, and $T$ is the corresponding temperature value.

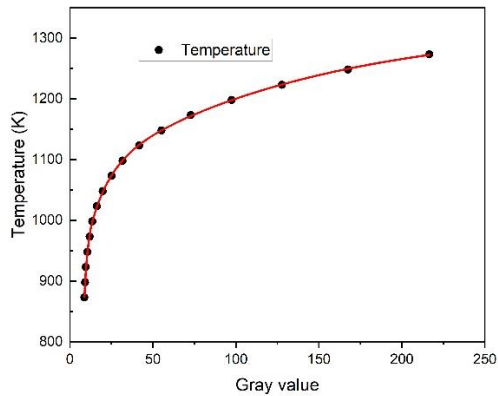

Fig. 14. Gray value-Temperature fitting curve at 768nm wavelength.

According to the relationship between grayscale and temperature in Formula 9, the grayscale images collected by 12 cameras are processed sequentially to obtain the butane flame's two-dimensional temperature projection in twelve directions. Figure 15(a) is the grayscale image of the flame captured by camera 1. Figure 15(b) shows the grayscale image's temperature map. The maximum temperature is 1230.78K.

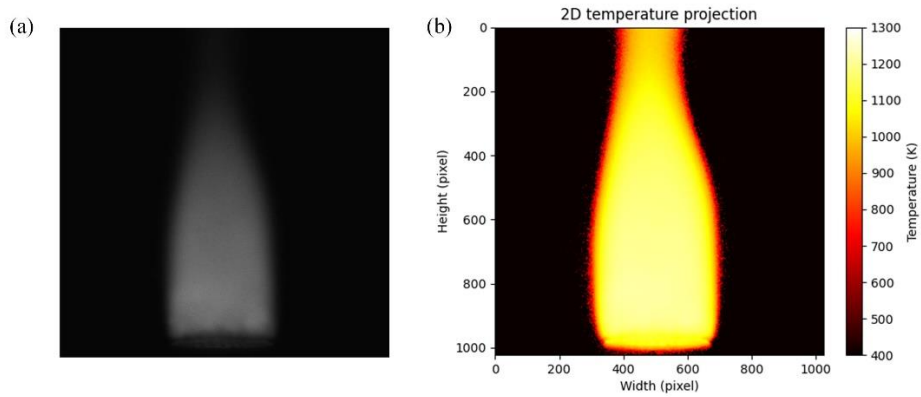

Fig. 15. The flame data collected by camera 1, figure (a) shows the flame grayscale map, and figure (b) shows the corresponding temperature map.

Figure 16 shows the change process of MSE loss value with epoch during SRRN training process. In this study, we regard all light rays completing one traversal as one epoch. As can be seen from the figure, after 20 epochs, the loss value gradually stabilizes at 7000.

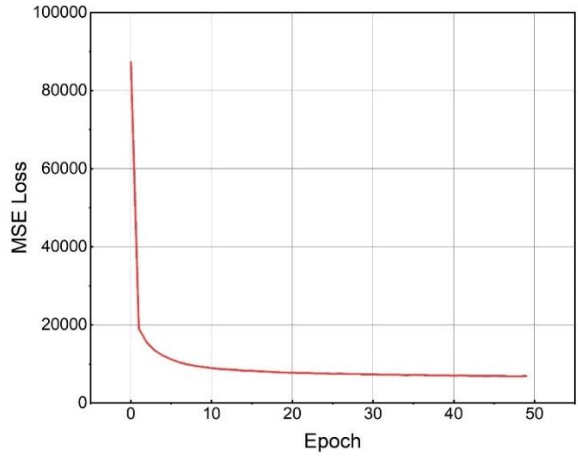

Fig. 16. Reconstruction of three-dimensional temperature field of butane flame with reprojection error.

In order to verify the accuracy of the butane flame's three-dimensional temperature field reconstruction results, we used K-type thermocouples to measure the temperature values at the three sampling points in Figure 17(a) and calculate the relative errors.

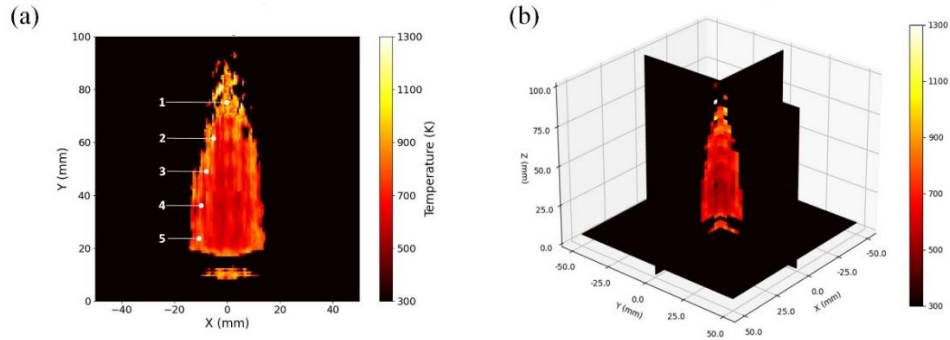

Fig. 17. Reconstruction of temperature field slices using butane flame. Figure (a) shows a two-dimensional slice of the reconstructed temperature field at x=0, points 1-5 are thermocouple sampling points, and figure (b) shows a three-dimensional slice of the reconstructed temperature field.

The results are shown in Table 1. Compared with the thermocouple measurement values, the reconstruction results have a minimum relative error of 4.69% and a maximum of 4.86%. As the sampling position increases, the flame temperature change trend is consistent with the thermocouple measurement value change trend. As shown in Table 2, SSRN has a better reconstruction effect at sampling point 1, while the reconstruction values at other positions are generally lower than the thermocouple measurement values. We think it may be due to the smaller flame volume at the flame tip. For the results in this study, The simplified thermo-optical radiation rendering formula can more accurately represent the relationship between the temperature field and its projection. It also points the way for us to improve the reconstruction algorithm in the next step.

Table 1. Verification of Radiation Temperature Measurement Results Based on Thermocouples

| Position | Thermocouple measurements (K) | Reconstructed value (K) | Relative error (%) |
|---|---|---|---|
| 1 | 1132 | 1077 | 4.86 |
| 2 | 895 | 853 | 4.69 |
| 3 | 724 | 689 | 4.83 |
| 4 | 707 | 673 | 4.81 |
| 5 | 685 | 652 | 4.82 |

## 5. Conclusion

This study proposes a deep learning-based SRRN model that uses a fully connected neural network to represent the three-dimensional temperature field implicitly. Compared with the traditional three-dimensional temperature field reconstruction technology that relies on voxels, this method can obtain the temperature value of that point by inputting the coordinates of any position in space and then obtaining a temperature field with any resolution.

In order to ensure the reliability of the light sampling results, we improved the differentiable rendering technology in computer graphics by studying the flame thermal and optical radiation model and proposed the thermal and optical radiation rendering technology. Since deep neural networks often suffer from vanishing and exploding gradients, we use residual connections to avoid this problem.

In order to improve the model's ability to represent rapidly changing areas within the temperature field, we use position encoding technology to map spatial coordinates to high-dimensional space, thereby converting high-frequency function fitting problems into low-frequency function fitting that neural networks are better at.

In the numerical simulation experiment, three temperature fields with different complexities were first reconstructed to verify the ability of the algorithm to represent temperature fields with different complexities. The reconstruction errors are 4.62 for the single-peak temperature field, 10.11 for the double-peak temperature field, and 10.17 for the three-peak temperature field. It can be seen from the results that SRRN also has a strong representation ability for complex temperature fields.

In the algorithm anti-noise experiment, Gaussian noise with 7% and 15% intensity was added to the projection data, respectively. The reconstruction results showed that SRRN has strong anti-Gaussian noise ability. However, the salt and pepper noise with an intensity of 7% significantly impacts the reconstruction results. We believe that the salt and pepper noise causes some light to be missing, which leads to incorrect reconstruction results at some locations in the space.

In the butane flame reconstruction experiment, 12 cameras were evenly arranged around the flame. Thermocouple points were used for verification, and the relative error was within 5%, indicating that SRRN has a solid ability to reconstruct the actual flame temperature field.
Although using the SRRN model as a function fitter to represent the three-dimensional temperature field solves the problems of low resolution of the traditional algebraic iterative reconstruction algorithm and the lack of real values in the supervised learning reconstruction, its reconstruction accuracy is highly dependent on the thermo-optical radiation model of the flame. It is more consistent to establish that an actual radiation model will be the focus of our subsequent research work.

**Funding.** National Natural Science Foundation of China (52075504); Shanxi Key Laboratory of Advanced Semiconductor Optoelectronic Devices and System Integration (No. 2023SZKF11); Shanxi Post graduate Education Innovation Program (2022Y627); Shanxi Post graduate Practical Innovation Program (2023SJ209); State Key Laboratory of Quantum Optics and Quantum Optics Devices (KF202301); Shanxi 1331 Project Key Subject Construction.

**Acknowledgments.** The authors thank all the reviewers, editors, and contributors for their contributions and suggestions as well as all the members of the OSEC Laboratory.

**Disclosures.** The authors declare no conflicts of interest.

**Data availability.** Data underlying the results presented in this paper are not publicly available at this time but may be obtained from the authors upon reasonable request.

**References**

1. Wang Z , Zhou W , Kamimoto T ,et al.Two-Dimensional Temperature Measurement in a High-Temperature and High-Pressure Combustor Using Computed Tomography Tunable Diode Laser Absorption Spectroscopy (CT-TDLAS) with a Wide-Scanning Laser at 1335–1375 nm:[J].Appl Spectrosc, 2020(2).DOI:10.1177/0003702819888214.
2. Rende W,Yingfei P,Xiaoyang T. Simultaneous measurement of temperature and water concentration using a novel laser dispersion spectrum extraction method immune to carrier phase variation[J]. Optics and Lasers in Engineering,2023,164.
3. Hao Z,Wanchen S,Liang G, et al. Optical diagnostic study of coal-to-liquid/butanol blend and dual-fuel combustion of a CI engine[J]. Fuel,2022,320.
4. DziarskiKrzysztofKrzysztof.Dziarski@put.poznan.plHulewiczArkadiuszInstitute of Electric Power Engineering,Poznan University of Technology,Piotrowo 3A, 60-965Poznan,PolandInstitute of Electrical Engineering and Electronics,Poznan University of Technology,Piotrowo 3A, 60-965Poznan,Poland.Uncertainty of Thermographic Temperature Measurement with an Additional close-up Lens[J].Measurement Science Review, 2021, 21(6):185-190.DOI:10.2478/msr-2021-0025.
5. Kong Q,Jiang G,Liu Y, et al. 3D high-quality temperature-field reconstruction method in furnace based on acoustic tomography[J]. Applied Thermal Engineering,2020,179(prepublish).
6. Haegele S , Corrielli G , Hejda M ,et al.Large field-of-view holographic imager with ultra-high phase sensitivity using multi-angle illumination[J].Optics and Lasers in Engineering, 2023.
7. Wang J , Oliveira P M D , Pathania R S ,et al.Stability and structure of lean swirling spray flames with various degrees of prevaporization:[J].International Journal of Spray and Combustion Dynamics, 2023, 15(2):91-104.DOI:10.1177/17568277231159173.
8. Rajendhar J , Ali S , Lasse L ,et al.Evaluating different deep learning models for efficient extraction of Raman signals from CARS spectra[J].Physical chemistry chemical physics: PCCP, 2023.
9. Senior W C B , Gejji R M , Gai T ,et al.Background suppression for CARS thermometry in highly luminous flames using an electro-optical shutter[J].Optics Letters, 2023.

10. Ge W , David C , Modest M F ,et al.Comparison of spherical harmonics method and discrete ordinates method for radiative transfer in a turbulent jet flame[J].Journal of Quantitative Spectroscopy & Radiative Transfer, 2023.
11. J. Floyd, P. Geipel, A.M. Kempf, Computed Tomography of Chemiluminescence (CTC): Instantaneous 3D measurements and Phantom studies of a turbulent opposed jet flame, Combust. Flame 158 (2011) 376–391.
12. Jin Y, Zhang W, Song Y, et al. Three-dimensional rapid flame chemiluminescence tomography via deep learning[J]. Optics express, 2019, 27(19): 27308-27334.
13. Cai W , Huang J , Deng A ,et al.Volumetric reconstruction for combustion diagnostics via transfer learning and semi-supervised learning with limited labels[J].Aerospace Science and Technology, 2021, 110:106487.DOI:10.1016/j.ast.2020.106487.
14. Remondino F, Karami A, Yan Z, et al. A Critical Analysis of NeRF-Based 3D Reconstruction[J]. Remote Sensing, 2023, 15(14): 3585.
15. Martinez I , Otamendi U , Olaizola I G ,et al.A novel method for error analysis in radiation thermometry with application to industrial furnaces[J].Measurement, 2022(2):110646.DOI:10.1016/j.measurement.2021.110646.
16. Hernando Maldonado Colmán, Cuoci A , Darabiha N ,et al.A virtual chemistry model for soot prediction in flames including radiative heat transfer[J].Combustion and Flame, 2022, 238:111879-.DOI:10.1016/j.combustflame.2021.111879.
17. Liwei C,Xianqi Z,Shan G, et al. Multi-spectral radiation thermometry based on an Alpha spectrum-LM algorithm under the background of high temperature and intense reflection.[J]. Optics express,2022,30(20).
18. Yang G , Yu Y , Sun Z ,et al.Radiometric calibration algorithm for high dynamic range infrared imaging system[J].Infrared physics and technology, 2023.
19. Fuhao Z,Hujie P,Xuan Z, et al. Three-dimensional reconstruction for flame chemiluminescence field using a calibration enhanced non-negative algebraic reconstruction technique[J]. Optics Communications,2022,520.
20. Yang Y, Hao X, Zhang L, et al. Application of scikit and keras libraries for the classification of iron ore data acquired by laser-induced breakdown spectroscopy (LIBS)[J]. Sensors, 2020, 20(5): 1393.
21. Liu X, Hao X, Xue B, et al. Two-dimensional flame temperature and emissivity distribution measurement based on element doping and energy spectrum analysis[J]. Ieee Access, 2020, 8: 200863-200874.
22. Bing X, Xiaojian H, Xuanda L, et al. Simulation of an NSGA-III based fireball inner-temperature-field reconstructive method[J]. IEEE Access, 2020, 8: 43908-43919.
23. I.A. M,J. Y,D. H, et al. Comprehensive characterization of sooting butane jet flames, Part 2: Temperature and soot particle size[J]. Combustion and Flame,2021(prepublish).
24. Schonberger J L, Frahm J M. Structure-from-motion revisited[C]//Proceedings of the IEEE conference on computer vision and pattern recognition. 2016: 4104-4113.